\documentclass[aps,pre,twocolumn,amssymb,showpacs,groupedaddress]{revtex4-2}
\usepackage{graphicx,color}
\usepackage[normalem]{ulem}
\usepackage{appendix}
\usepackage{hyperref}
\usepackage{amsmath}
\usepackage{tabularx}
\usepackage{comment}
\def\beq{\begin{equation}}
\def\eeq{\end{equation}}
\def\bea{\begin{eqnarray}}
\def\eea{\end{eqnarray}}

\def\cL{\mathcal{L}}

\def\cW{\mathcal{W}}

\def\q0{q_0}

\begin{document}
\definecolor{colortodo}{RGB}{255,0,0}

\newcommand{\red}[1]{{\color{colortodo}#1}}
\title{Bundling architecture in elastic filaments with applied twist}

\author{Amit Dawadi$^\dagger$, Animesh Biswas$^\dagger$, Julien Chopin$^{\star}$, and Arshad Kudrolli$^\dagger$}
\affiliation{$^\dagger$ Department of Physics, Clark University, Worcester, MA 01610, USA\\
$^{\star}$ Instituto de F\'isica,  Universidade Federal da Bahia, Salvador-BA 40170-115, Brazil}
\date{\today}

\begin{abstract}
We investigate the formation of helical multifilament bundles and the torque required to achieve them as a function of applied twist. Hyperelastic filaments with circular cross sections are mounted parallel in a uniform circle onto end-clamps that can move along the twist axis depending on the applied axial load. With increasing twist, the filaments describe a hyperbolic hyperboloid surface before coming into contact in a circle, and then packing in a tight helical bundle in the center with increasing twist. While the bundle appears ordered for sufficiently small number of filaments, they are disordered for large enough number of filaments and applied twist. We reveal with x-ray tomography, that the packing of the filaments becomes disordered following a radial-instability which leads to a decrease in bundle radius, and migration of filaments relative to each other in the bundle. Nonetheless, the helical angle of the filaments in the bundle are found to be essentially constant, resulting in inclination angles which increase with distance from axis of rotation. We develop energy minimization analysis to capture the observed variations in bundle length and torque as a function of number of filaments considering the neo-Hookean nature of the filaments.  We show that the bundle geometry and the applied load can be used to describe the non-linear torque profile measured as a function of twist angle.  
\end{abstract}

\maketitle
\section{Introduction}
Twisted filaments can be found widely in organic and synthetic matter going back to antiquity~\cite{Bohr2011}. They are an important element in fabricating metamaterials such as artificial spider silk with exceptional tensile strength, flexibility, and toughness~\cite{Dou2019, Zhou2021}. Slender fibers are spun together to not only gather them, but also increase their collective strength in making yarns as twisting increases the friction and mechanical interlocking between fibers~\cite{Seguin2022,Hanlan2023,Crassous2024}. However, making optimal yarns requires a deep understanding of their kinetics as it has been shown that yarn strength increases and then decreases with twist as competing factors cohesion and fiber inclination angle become dominant in the low and high twist regimes, respectively~\cite{Schwartz2019}. In the biological realm, collagen fibrils 
which bind muscle with bones through tendons and provide structural and mechanical support have been shown to have rope-like structure~\cite{Bozec2007}, and bundling and unbundling of flagellar filaments is important for the locomotion of Peritrichous bacteria~\cite{Darnton2007}. 
Helical twist plays a role in stabilizing sickle cell disease causing sickle-hemoglobin macrofibers by determining their structural arrangement and energy landscape~\cite{Makowski1986, Turner2003}.  
Thus, functionalities of twisted filaments in biological systems~\cite{Mitchison1996, Leijnse2022}, besides emerging applications such as in twistron energy harvesters~\cite{HyeongKim2017}, high 
efficiency twistocaloric coolers~\cite{Wang2019}, and actuators in soft robotics~\cite{Shoham2004, Lima2012, Haines2014,Mahadevan2019, Kanik2019, Wang2022, Lamuta2023}, make them an important system to study. 

While linear filaments with uniform circular cross sections can bundle tightly with hexagonal order when stacked together, it has long been realized that the packing grows disordered when they are twisted collectively since the filaments assume a helical shape with increasing curvatures as the filaments push out against each other~\cite{Grason2015}. The filament transects orthogonal to the axis of rotation become non-circular with increasing distance from axis of rotation and they can no longer close pack with the same hexagonal symmetry.  Thus, the twisted packing structure of ropes and cables that optimize their strength and volume have been a subject of patents for well over a century. A mapping of the bundle cross section to a packing of disks on a non-Euclidean surface was introduced to describe the overall structure of the bundle with twist~\cite{Bruss2013}, and was tested with bendable but otherwise inextensible filaments~\cite{Panaitescu2017}. 

However, elasticity plays a crucial role in the behavior of twisted filament bundles~\cite{Atkinson2021}, affecting their order persistence~\cite{Panaitescu2018} and promoting bundling through configurational instabilities and hydrodynamic interactions~\cite{YMan2017}. The effect of elasticity was recently investigated by measuring the torque required to twist hyperelastic filament pairs and comparing it with energy-based models~\cite{Chopin2024}. Building on this research, we examine the packing, torque, and stored energy in multifilament bundles created by twisting initially parallel filaments that are initially held apart under constant tension. In contrast with previous studies~\cite{Panaitescu2017,Panaitescu2018}, we measure the torque required for bundle formation and present a combined geometric and energy-based model to describe bundle length and torque as a function of applied twist enabled by x-ray measurements of their internal structure.

The paper is organized as follows. We first discuss the experimental apparatus used to twist the filaments and measure its response using optical imaging and a torque sensor. Then, we discuss x-ray scanning to obtain and characterize the internal structure of the bundle. Based on these observations, we analyze the trends in the observed torque as a function twist angle and the evolving geometric parameters important to determining its magnitude. Then, we develop an elastogeometric model to explain overall growth of bundle length, energy stored, and torque with applied twist.

\section{Experimental System}
\begin{figure}[h]
\begin{center}
\includegraphics[width=0.45\textwidth]{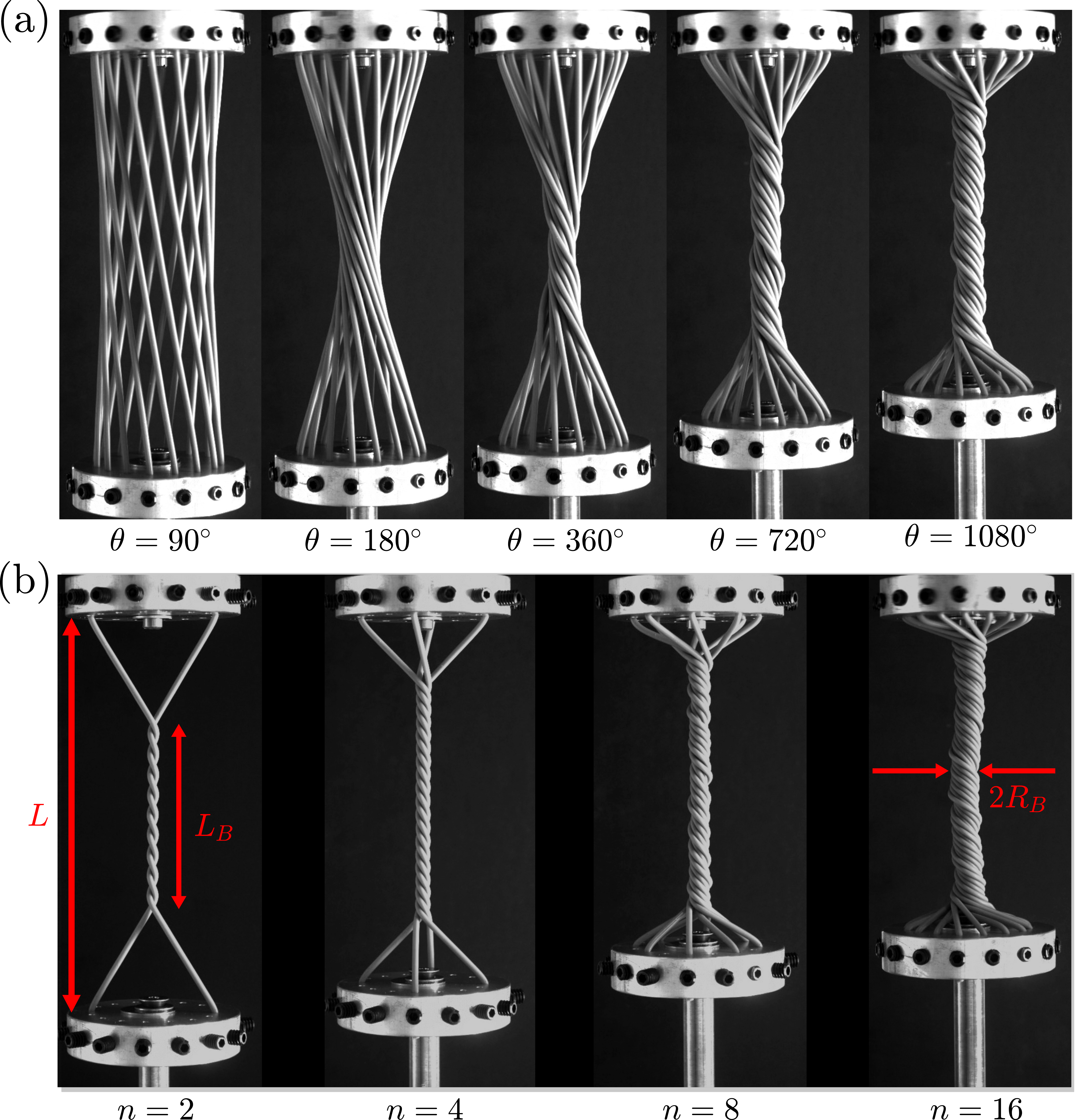}
\end{center}
	\caption{(a) Formation of filament bundle with increasing applied twist angle $\theta$ ($L_f = 162$\,mm; $n=16$). Before filaments come in contact and form a bundle, they extend linearly between their clamped ends and thus fall on a hyperbolic hyperboloid surface.
    (b) The bundles observed with increasing number of filaments ($L_f = 162$\,mm; $\theta = 1800^\circ$). The end to end length $L$, bundle length $L_B$ and radius $R_B$ used to characterize the evolving geometry of the system are indicated. 
    } 
	\label{fig:filaimages}
\end{figure}

The experiments are performed with silicone filaments obtained from MSC Industrial Supply that have circular cross sections with diameter $d_0=2.60$\,mm, Poisson ratio $\nu \approx 0.5$, and shear modulus $\mu = 1.43 \pm 0.1$\,MPa~\cite{Chopin2024}. Their bundling mechanics is observed while varying the filament number $n$ between 2 and 16.
The filaments are mounted in parallel onto two circular clamps with radius $R_0  = 25$\,mm and diameter $D_0 = 50$\,mm separated equally by angle $360^0$ divided by $n$ such that the relaxed length of the filaments between their clamped ends $L_f = 162$\,mm. 
The filaments are twisted by rotating one of the clamped ends with a Parker Motion Control system stepper motor through prescribed angle $\theta$ about the axis joining the centers of the two ends. The other clamped end is allowed to move along the twist axis with linear guides so that the distance between the clamps $L$ can vary in response to the applied load $F$ and the applied twist angle $\theta$.  A load proportional to the number of filaments is applied at the moving end to taut the filaments, resulting in a total force $F = n F_1$
, where $F_1 = 0.735$\,N. In practice we find that an additional frictional force $F_\mu$ associated with the linear guides can contribute to the axial load which is estimated to be approximately 1\,N. 
The applied load along the symmetry axis results in an increase in filament length $\Delta L = \frac{2 F L_f}{n\pi \mu(1+\nu) d_0^2 } \approx 5.5$\,mm. 
Thus, the total clamped end to end length $L_0 = L_f+ \Delta L \simeq  168$\,mm, corresponding to a strain $\Delta L/L_f = 0.034$ before application of twist.

\begin{figure}[h]
\includegraphics[width=0.4\textwidth]{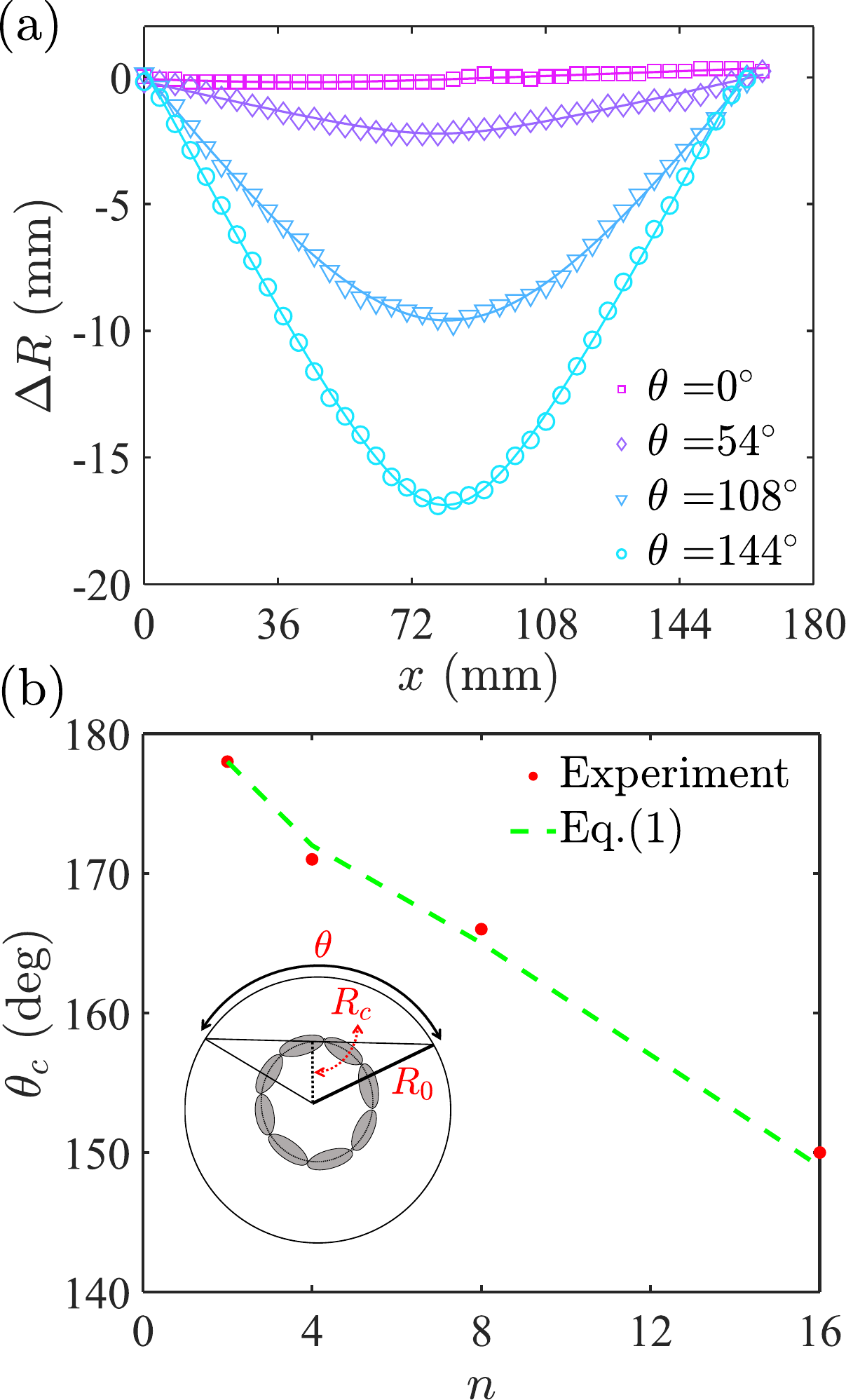}
	\caption{(a) The tracked bundle edge as viewed orthogonal to the axis of twist compared with a hyperbolic function. Its surface of revolution about the axis of twist corresponds to a hyperboloid. (b) The estimated and measured $\theta_c$ as a function of $n$ which separates the hyperboloid and bundling regime ($L_f = 162$\,mm). Inset: A schematic of the filaments geometry as viewed along the twist axis for $\theta = \theta_c$ when filaments just come into contact. The filament cross sections are elliptical because of their inclination relative to the twist axis. 
    } 
	\label{fig:schem}
\end{figure}

Side view images obtained with a Pixelink camera are shown at various $\theta$ between $90^0$ and $1080^0$ for $n=16$ in Fig.~\ref{fig:filaimages}(a), and  for $\theta = 1800^0$ as $n$ is increased from 2 to 16 in Fig.~\ref{fig:filaimages}(b). 
We observe that the filaments tilt with respect to the axis of rotation, move inwards until they come in contact at an angle $\theta_c$ before $\theta$ reaches $180^0$, and then wrap around each other in the center as $\theta$ is increased further. One also observes that the clamp end to end distance decreases with increasing twist and number of filaments. The filaments appear straight away from the clamped ends before contact and can be then expected to fall on a ruled surface which is a hyperboloid of one sheet~\cite{hilbert1952geometry}. Therefore the projected boundary when viewed perpendicular to the axis of twist is a hyperbola as shown in Fig.~\ref{fig:schem}(a) before filaments come into contact for $n=16$ case.  After contact, the filaments form a bundle at the center which grows towards the clamped ends with increasing twist. The bundles appear ordered in the case of $n=2$ and $4$, but disordered when $n=8$ and $n=16$ when $\theta = 360^0$ or higher in the examples shown. Further, incipient longitudinal buckling  can be noted in the image corresponding to $n=16$ at $\theta = 1800^0$ as the bundle ends approaches the clamped ends. We confine our discussion to the bundling regime before the development of longitudinal buckling in this study. 

\begin{figure*}
\includegraphics[width=0.95\textwidth]{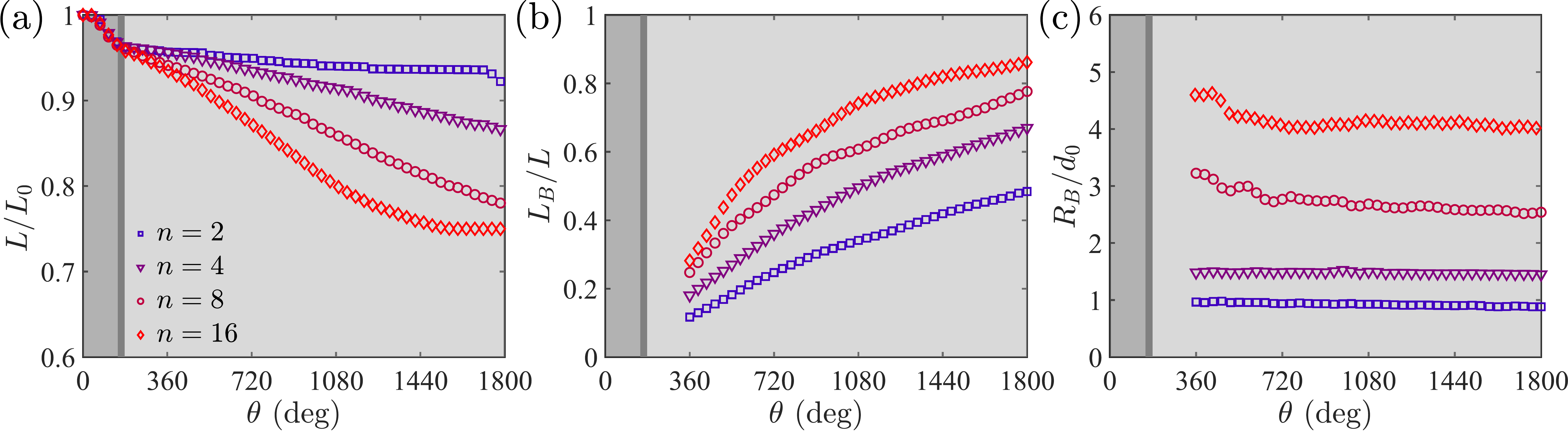}
	\caption{(a) The clamp end to end distance $L$ scaled by the pre-twist length $L_0$ as a function of $\theta$ ($L_0 = 168$\,mm). The rate of decrease of $L$ is observed to change as the filaments come into contact around  $\theta_c$ which is indicated by a vertical line. (b) The bundle length $L_B$ increases after filaments come in contact and approaches $L$ with increasing $\theta$. (c) The bundle radius $R_B$ scaled by filament diameter $d_0$ is essentially constant with increasing $\theta$.}
	\label{fig:filageom}
\end{figure*} 
We can estimate $\theta_c$ which separates the observed hyperboloid and bundle regimes by noting that the filaments extend essentially linearly between the clamped ends, and come in contact in a circle with a radius $R_c$ by symmetry (see Inset to Fig.~\ref{fig:schem}). 
Then, $\cos(\frac{\theta_c}{2}) = \frac{R_c}{R_0}$. The filaments are inclined with angle $\alpha_f$  w.r.t the twist axis when they come in contact. Nonetheless, in the case of $n=2$, the distance between filament centers is simply $d$, where $d$ is the filament diameter which is less than $d_0$ due to the stretching of the filament length due to Poisson effect. However, as illustrated in Inset to Fig.~\ref{fig:schem}(b), the filaments touch along the long axis of their elliptical cross sections for large enough $n$ in the plane perpendicular to the twist axis. Then, the distance between contacts with neighbors is given by $d/\cos{\alpha_f}$. Thus, $2\pi R_c = nd/\cos{\alpha_f}$. One can note that the filaments do not stretch significantly further while twisted by $\theta_c$ after application of the axial load, i.e. $\cL \approx L_f + \Delta L = 168$\,mm, $d = d_0(1 - \nu \Delta L /L_f) \approx 2.5$\,mm, and $\theta_c$ is close to $180^0$ when the filaments come in contact \footnote{The increase in tension of the filament by a factor $\sec{\alpha_f}$ due to the inclination of the filaments result in change of filament length of $5.76$\,mm, instead of $\Delta L = 5.5$\,mm, and thus has negligible effect on the estimated $\theta_c$.}. Then, $\sin{\alpha_f} \approx 2 R_0/(L_f + \Delta L)$. 
Thus, 
\begin{equation}
\begin{split}
   \cos(\frac{\theta_c}{2}) &\approx \frac{d}{2 \pi R_0}\,,  
   \,\, {\rm for}\,\, n = 2, \,\, {\rm and} \\
   \cos(\frac{\theta_c}{2}) & \approx \frac{nd}{2\pi R_0 \sqrt{1-(\frac{2 R_0}{L_f + \Delta L})^2}}\,, 
   \,\, {\rm for}\,\, n > 2.
    \label{eq:thetac}
\end{split}
\end{equation}
The estimates for the various $n$ are plotted in Fig.~\ref{fig:schem}. These estimated values are consistent with those obtained by visual inspection. We measure $\theta_c$ more accurately using torque measurements as we discuss later in Section~\ref{sec:torque}.

\section{Filament assembly}

We quantify the overall evolution of the filament assembly in terms of the end to end distance between the clamps $L$, the bundle length $L_B$ and the bundle radius $R_B$ (see Fig.~\ref{fig:filaimages}). We obtain $L$ by identifying the clamp edges from the images and plot it scaled by $L_0$ in Fig.~\ref{fig:filageom}(a) as a function of $\theta$ for various $n$. The range of $\theta_c$ according to Eq.~(\ref{eq:thetac}) are indicated for reference. Before contact,  $L/L_0$  can be observed to overlap and decrease with $\theta$. Whereas, $L/L_0$ can be observed to decrease at different rates depending on the number of filaments after contact. Although not quite as fast as before contact, the overall decrease can be observed to be faster for increasing $n$ when $\theta > \theta_c$.  For inextensible filaments, the decrease in distance is a consequence of the evolving geometry alone. However in the case of elastic filaments, a detailed calculation is necessary which we present in Section~\ref{sec:model}.    

While the bundle may be considered to form, starting from when the filaments first come in contact at $\theta = \theta_c$, the bundle length itself is difficult to measure from the side view images. Therefore, we plot the bundle length $L_B$ in Fig.~\ref{fig:filageom}(b) for $\theta > 360^0$ when we can determine the length clearly by processing the images.  We observe that $L_B$ increases rapidly after the filaments come in contact but slows down as it increases towards $L$.  The bundle lengths for the same $\theta$ are systematically higher for greater $n$ showing that the filaments not only wrap around each other but also push laterally against each other leading to a relatively greater $L_B$. 

Complementarily, the bundle radius $R_B$ can be also measured from the projection of the bundle cross section on the image plane after the filaments have wrapped around at least once. Fig.~\ref{fig:filageom}(c) shows that $R_B$ increases systematically with $n$. 
It can be also noted that $R_B$ remains more or less constant, if not decreases slightly, for $\theta > 360^0$, when it can be cleanly identified. Because filaments in the bundles incline with increasing twist, one may expect the radius to increase~\cite{Bruss2013}. However, the filaments can stretch as the tension in the filaments increase with twist, which can lead the radius of individual filaments to decrease. The observation that the bundle radius remains essentially constant suggests that these two opposing effect appear to cancel each other in filament bundles formed under constant applied load conditions.

\section{Bundle structure}

\subsection{X-ray tomography}
\label{sec:xray}
\begin{figure}
\includegraphics[width=0.45\textwidth]{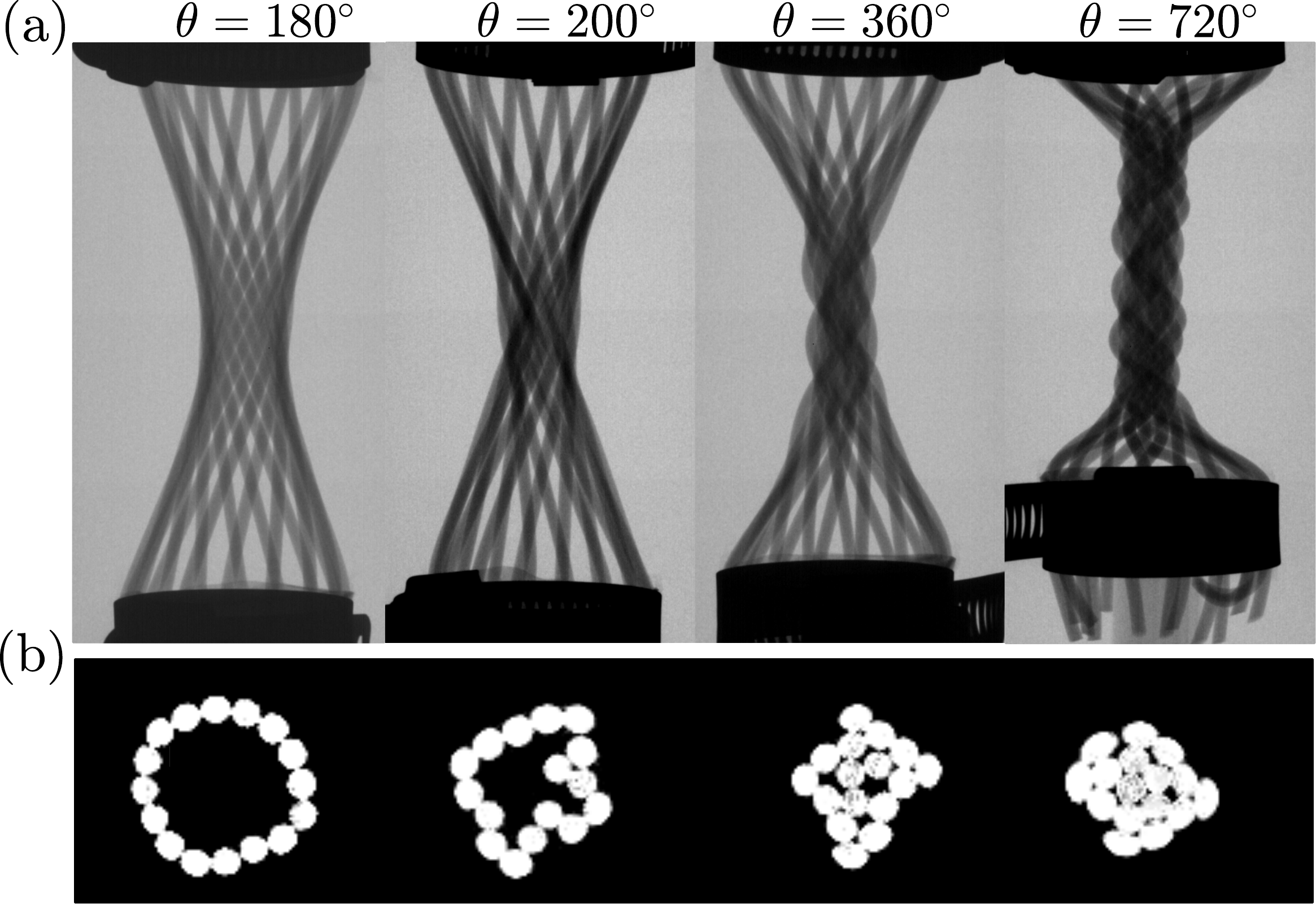}
\caption{(a) Radiograms of the bundle with increasing $\theta$. (b) The corresponding central cross-sections obtained with Computed Tomography. The filaments start to come into contact in a ring at approximately $\theta_c = 135^0$, and filaments undergo radial instability at $\theta \simeq 200^0$. The filaments form a tight disordered bundle which grows in length with increasing twist ($L_0 = 11$\,cm; $D_0 = 4$\,cm).   }
	\label{fig:radio}
\end{figure} 

\begin{figure}
\includegraphics[width=0.4\textwidth]{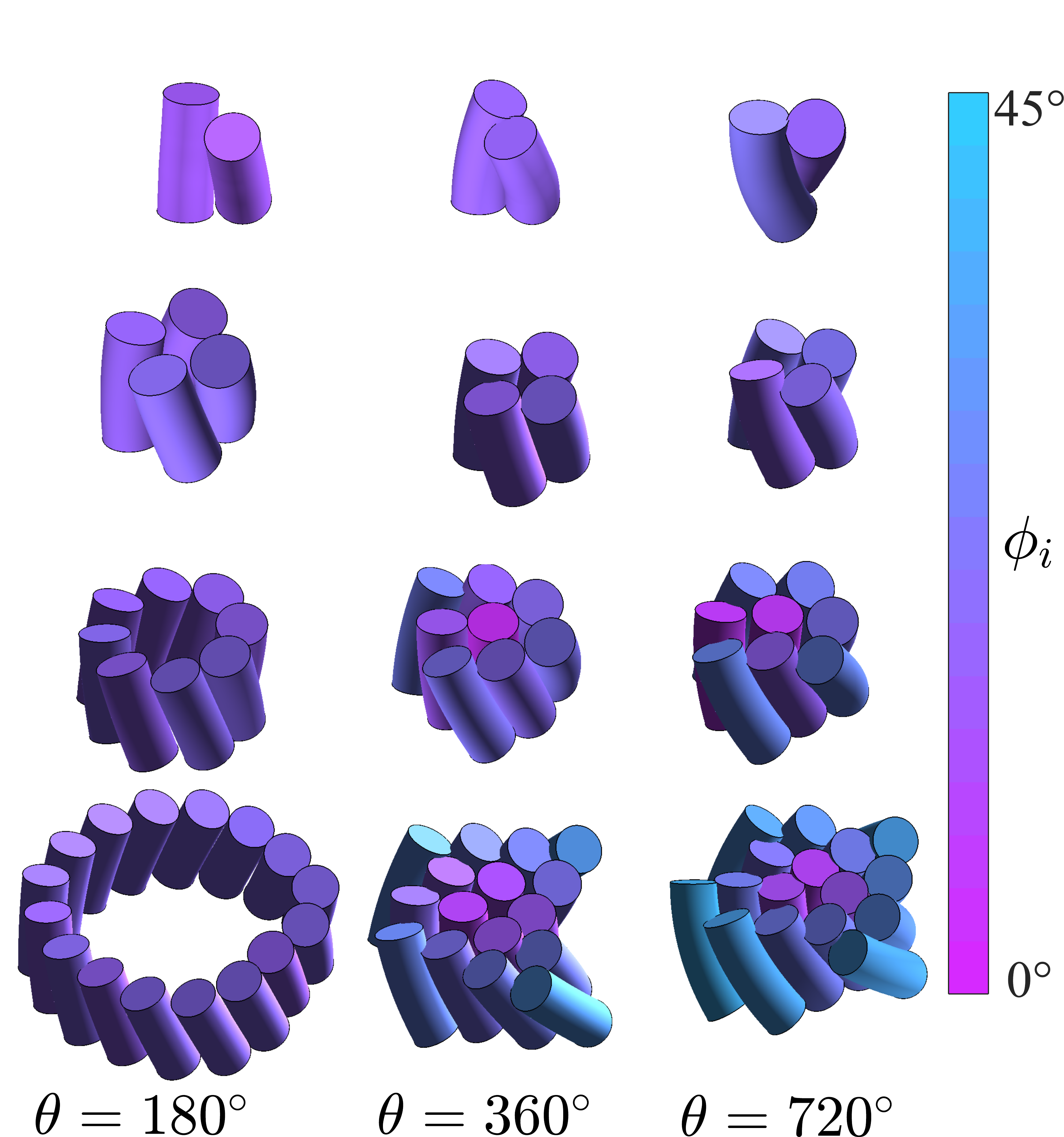}
\caption{The filament bundles ($n=2, 4, 8, 16$) tracked in a central section of the bundle at $\theta$ = $180^\circ$, $360^\circ$, and $720^\circ$ with inclinations according to the color map. The filaments initially come in contact in a ring because they are mounted symmetrically at their ends on circular clamps and then make tight bundle after radial instability with further twist.  The filaments have the same diameters but shown with decreasing magnification with $n$.}
	\label{fig:x-bundle}
\end{figure}

We perform x-ray scans to further reveal the internal structure of the filaments, and their arrangements at different cross-sections as they twist around each other and form the bundle. To fit the apparatus in the Varian BIR 150/130 Desktop CT system machine, we constructed a scaled down version of the apparatus with $R_0 = 20$\,mm and $L_f = 110$\,mm to apply twist while using the same silicone filaments. Figure~\ref{fig:radio} shows a set of radiograms and central bundle cross sections in the case of $n=16$ as the bundles begin to form, after they buckle, and as the bundle grows and fills the entire length $L$ with twist. We observe that the filaments are arranged symmetrically at $\theta = 180^0$ well after coming in contact at $\theta_c \approx 140^0$. As $\theta$ is increased further to $200^0$, the filaments are no longer arranged symmetrically. 

Fig.~\ref{fig:x-bundle} shows the rendering of an approximately 5\,mm mid-section of the bundle for the different $n$ with increasing $\theta$.  A similar order to disorder progression is observed for $n=8$. However, the filaments remain ordered as they twist around each other in case of $n=2$ and $n=4$, and no rearrangements are observed over the range of twist angles investigated. Thus, while $\theta_c$ estimated using Eq.~(\ref{eq:thetac}) are systematically lower for these bundles formed with $R_0 =20$\,mm versus the ones formed with $R_0 = 25$\,mm, the overall development of the bundles are observed to be similar in scaled down version. 

\begin{figure}
\includegraphics[width=1\linewidth]{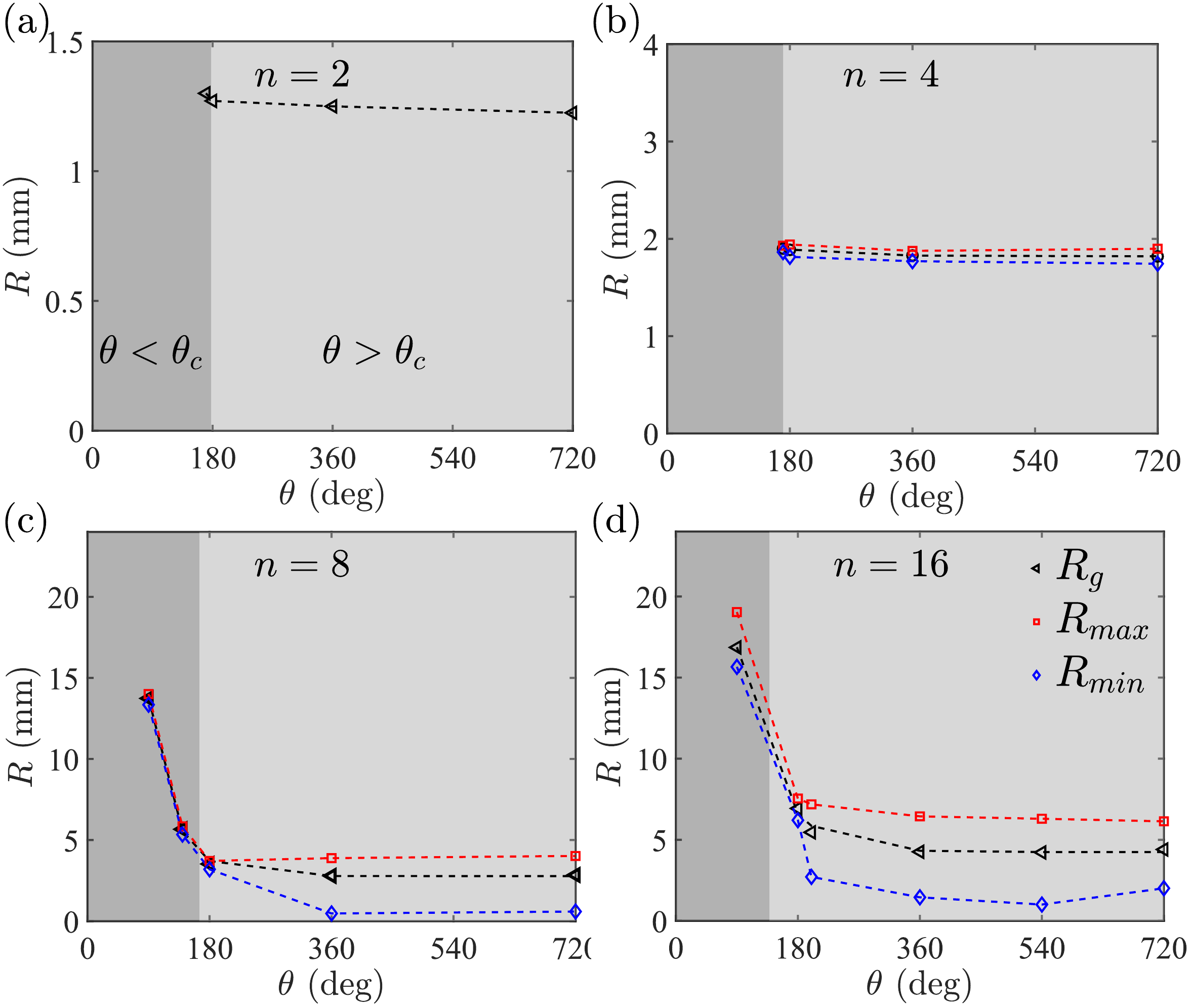}
	\caption{$R_g$, $R_{min}$ and $R_{max}$ obtained from the filament positions tracked with x-ray tomography as a function of applied twist ($R_0 = 20$\,mm; $L_0=110$\,mm) for $n=2,4, 8,$ and $ 16$. } 
	\label{fig:rbundle}
\end{figure}

We locate the centers of each of the filaments in the central cross section, and obtain the radius of gyration $R_g$, the distance to the closest filament $R_{min}$, and the farthest filament $R_{max}$ from the bundle center. These are plotted for $n=2, 4, 8,$ and 16 in Fig.~\ref{fig:rbundle}.  In the case of $n=2$, $R_{min}$, $R_{max}$, and $R_g$ are the same by definition. Further, they are also the same within filament mounting errors in the case of $n=4$. This shows that they remain packed at the corners of a square, and do not undergo radial instability over the range of $\theta$ studied. However, $R_g$, $R_{min}$ and $R_{max}$ differ after the radial instability occurs and some of the filaments move towards the center in the case of $n=8$ and 16. Just as we observed with optical imaging, we find that $R_{max}$ remain essentially constant as the bundle length grows. Here, we further observe that and $R_g$ remains essentially constant as well with increasing twist.  

\subsection{Bundle helicity}
\label{sec:helicity}
\begin{figure}
\includegraphics[width=8cm]{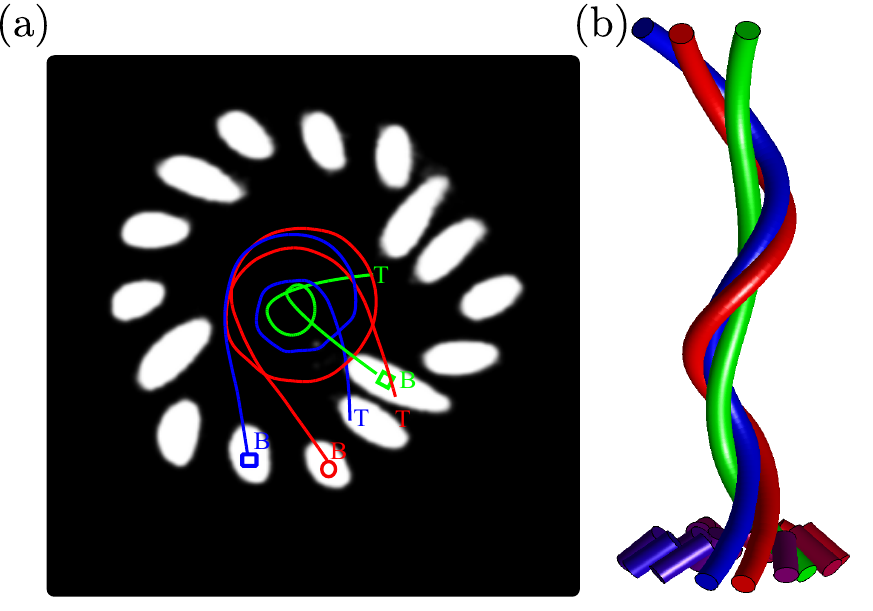}
\caption{(a) A cross section of the filaments in the fan out region between the clamped ends and the bundle ($n=16$). The filament cross sections appear oblate because of their inclination. The projected position of the three representative filaments are also traced as viewed along the twist axis. The positions are obtained between the top and the bottom of the bundle shown in Fig.~\ref{fig:radio}. (b) The tracked rendering of the three filaments. A short section of the other filaments are also rendered to indicate their initial position. The filaments can be noted to form helicoidal wrappings, but their distance from the center is observed to vary indicating that filaments migrate in the bundle as they wrap around the twist axis.}
\label{fig:bundle_view}
\end{figure}

Figure~\ref{fig:bundle_view}(a) shows a cross section of the filaments in the fan out region between the bundle and clamped ends. While the filaments have circular cross sections, the cross sections in the slice perpendicular to the axis of rotation appear oblate because of their inclination relative to the twist axis. While all the filaments are clamped at their ends at a distance $R_0$ from the axis around which twist is applied, they enter the bundle at different distances from the twist axis after the development of radial instability. Thus, their inclination angle with respect to the axis of twist vary, and the ones which enters closer to the axis of twist appearing more oblate.  The tracked centroid of the representative filaments are marked, and their projected positions as they twist around in the bundle and fan out at the other end are shown as continuous lines. A three-dimensional rendering of the tracked filaments is shown in Fig.~\ref{fig:bundle_view}(b), and an animated tracking of the filaments can be found in the supplementary documentation~\cite{sup-doc}. It can be seen from these tracked positions that the filaments do not maintain a constant helical radius along the entire length of the bundle.   Thus, while the filaments appear to have an overall helical shape, they in fact  migrate slowly in and out relative to each other from one end of the bundle to the other.  

The migration in the filaments is observed in the bundles with $n=8$ as well after a radial instability occurs. Thus, migration occurs when the filaments are brought together from some distance apart and then twisted around each other following a radial instability which leads to disordered packing. 
The relative migration of filaments has been noted in production of yarns~\cite{HearleBook, Treloar1965, Hearle1965}, but was not observed in previous studies when the filaments were hexagonally packed initially, and then twisted~\cite{Panaitescu2017,Panaitescu2018}.  The migration of the filaments can mitigate the dispersion in their extension, relieving the compression which can otherwise develop in the central filaments that are not as stretched as those at the edges. Thus, the method by which the bundles are prepared have a significant effect on the bundle structure and the development of disorder. 

\begin{figure}
    \centering
    \includegraphics[width=0.95\linewidth]{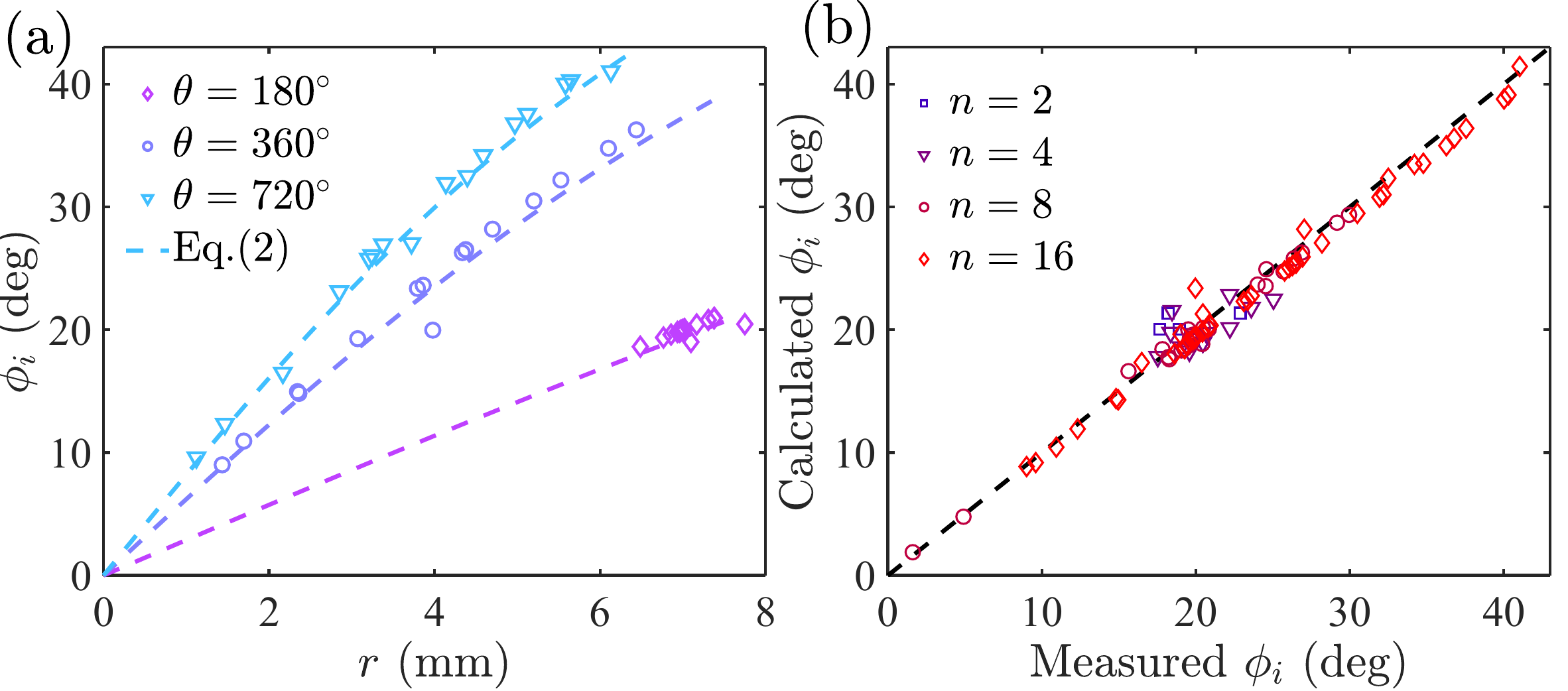}
    \caption{(a) The angle of inclination of filaments $\phi_i$ as a function of distance from the axis of rotation $r$ for $n=16$ and $\theta = 360^0$ and $720^0$.  Curves corresponding to Eq.~(\ref{eq:phi_f}) are also plotted and observed to be good agreement with the data. (b) The measured and calculated $\phi_i$ using Eq.~(\ref{eq:phi_f}) for various $\theta$ and $n$ are also in agreement as they fall close to the dashed line with unit slope. Filament migration leads to small deviations from the overall trends.}
    \label{fig:phi_f_rho}
\end{figure}

\subsection{Bundle region filament inclinations}
We measure the angle of inclination of the filaments in the bundle $\phi_f$ relative to the axis of rotation from the tracked position of the filaments at the central cross section.  The filaments are rendered in Fig.~\ref{fig:x-bundle} corresponding to increasing applied twist angles {blue for all $n$} as they first come in contact in a uniform circle and then pack in increasingly tightly. From the color bar it is clear that the filaments are increasingly inclined with increasing distance from the twist axis. We plot $\phi_i$ corresponding each filament $i$ in Fig.~\ref{fig:phi_f_rho}(a) for three different $\theta$ in the 16-filament bundle. We observe that $\phi_i$ are clustered together when $\theta = 180^0$ in a tight circle with a small dispersion in distance from the bundle center $r$. The range of distances $r$ can be observed to decrease after the radial instability. 

If a filament has a helical shape, then the filament tilt angle w.r.t. axis of rotation  $\phi_i$ is given by  $\phi_i = \tan^{-1}\Big(\Omega_B  \, r_i \Big)$, where $\Omega_B$ is its helical angle~\cite{Panaitescu2018} and $r_i$ is the distance to the $i^{th}$  filament from axis  in the helical bundle. We estimate $\Omega_B$ by measuring the twist angle of the filament in the bundle $\Delta \theta = \theta - \theta_c$ and $L_B$, i.e. $\Omega_B = (\theta-\theta_c)/L_B$. Substituting, we get 
    \begin{equation}
    \tan{\phi_i} = \frac{(\theta-\theta_c)}{L_B}  \, r_i. 
    \label{eq:phi_f}
\end{equation}
The inclination angle $\phi_i$ obtained from Eq.~(\ref{eq:phi_f}) are shown in Fig.~\ref{fig:phi_f_rho}(a) as dotted lines for the three different twist angles after contact. The error in determining $L_B$ leads to some dispersion in estimating $\phi_i$, but these are found to be less than the marker size. A comparison of measured and estimated $\phi_i$ for each filament with Eq.~(\ref{eq:phi_f}) are shown in Fig.~\ref{fig:phi_f_rho}(b) corresponding to various combinations of $n$ and $\theta$. Good overall agreement is observed which tells us that the inclination angle of the filaments in the bundle are mostly determined by their distance from the axis of the bundle and the overall twist applied to the bundle. 

As we noted in the discussion of Fig.~\ref{fig:bundle_view}, the filaments migrate as they move from one end of the bundle to the other and the helices they form do not have a constant radii. This fact leads to some dispersion, but apparently the effect is small according to our measurements compared with the overall increase in $\phi_f$ with distance from the central twist axis.

\subsection{Fan out region filament inclinations}

\begin{figure}
    \centering
    \includegraphics[width=1\linewidth]{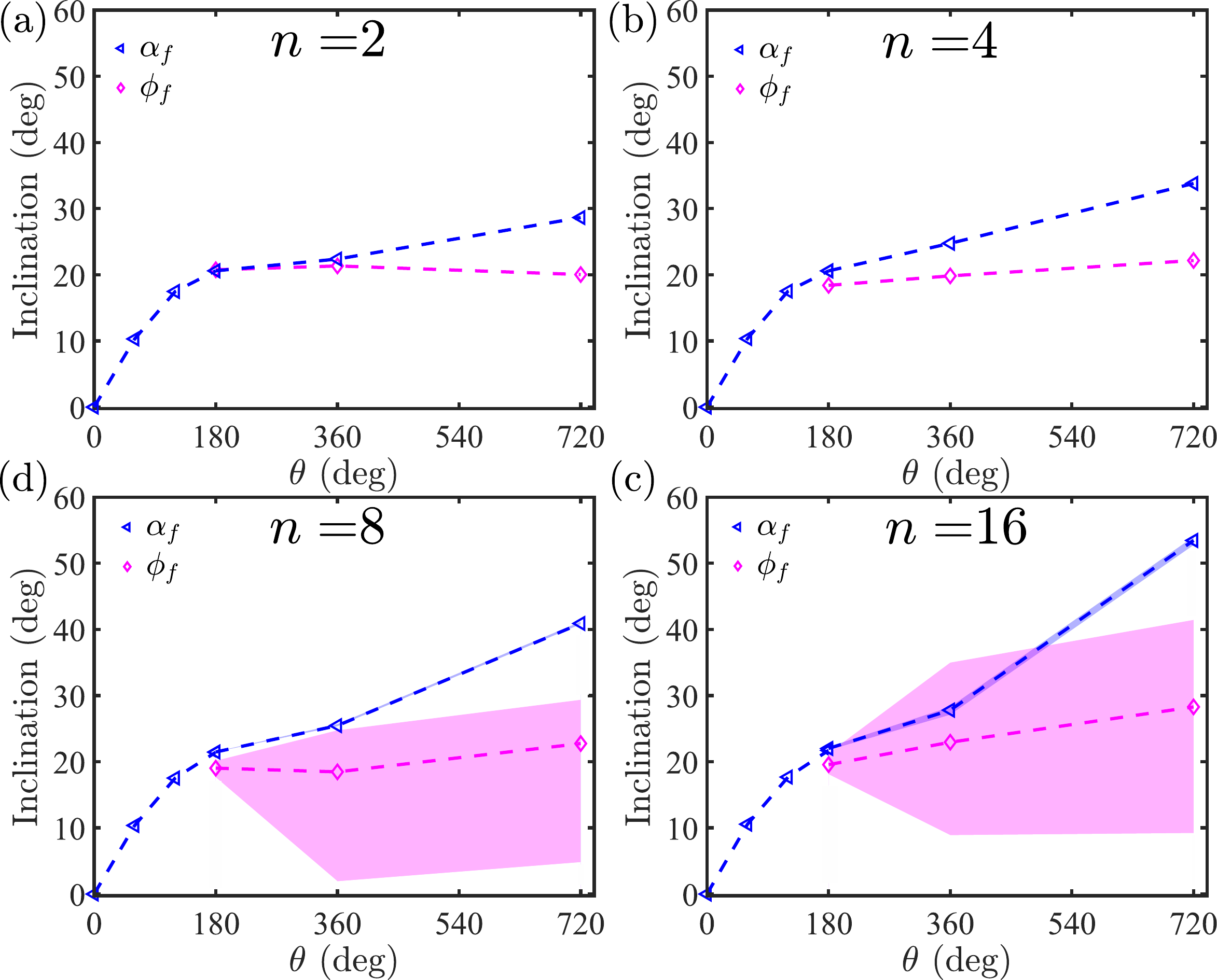}
   \caption{ (a-d) The average angle of inclination of the filaments in the fan out region  ($\alpha_f$) and in the bundle region ($\phi_f$) obtained from x-ray CT measurements. Inclination of filaments inside the bundle and in the fan out region fall respectively inside the shaded regions with magenta and blue colors.}
   \label{fig:x_ray_inclination1}
    \end{figure}

We examine the inclination angles of the filaments in the fan out region and their relation to the inclination angles of the filaments in the bundle. Before the formation of the bundle, we have $\alpha_f =  \tan^{-1}{(2R_0\sin(\theta/2)/L)}$ assuming that the filaments are uniformly arranged and clamped in a circle. If a filament $i$ enters the bundle at the center, the filament inclination angle $\alpha_i = \tan^{-1} \big( \frac{2R_0}{L -L_B}\big)$. For the filaments corresponding to $r_i$, the angle would be smaller and can be estimated as $\alpha_i = \tan^{-1} \big( \frac{2\sqrt{R_0^2 -r_i^2}}{L -L_B}\big)$. Although $\alpha_i$ for the filaments is not the same
because of the differences in where the filament enters the bundle, we find that in practice $\alpha_i$ are in fact fairly narrowly distributed around their mean value $\alpha_f$. I.e, the dispersion in $\alpha_i$ is small for $r_i \ll R_0$, and we only consider their mean value $\alpha_f$. Fig.~\ref{fig:x_ray_inclination1} shows a plot of measured angles $\alpha_f$ which shows that it increases rapidly before contact and then less rapidly for $\theta > \theta_c$.  

Considering that $r_i$ are the same in the case of $n=2$ and 4, we plot their average $\phi_f$ as a function of $\theta$ in Fig.~\ref{fig:x_ray_inclination1}(a) and Fig.~\ref{fig:x_ray_inclination1}(b), respectively. In case of $n=8$ and 16, we denote the range of $\phi_i$ which is bounded above corresponding to the outermost layer, i.e., $r_i = R_{max}$, and below by a filament which corresponds to $R_{min}$. We observe that while angle of the inclinations of the filament in the bundle region starts out at a value which is similar to $\alpha_f$, it grows relatively slowly with increasing $\theta$ in comparison to $\alpha_f$. Further, $\phi_i$ are widely dispersed as indicated by the shaded regions for compact bundles which have undergone a radial instability. Overall we observe that while filament inclination angles in the bundle region and in the fan out region start out being similar, the inclinations in the fan out region grow faster compared with those in the bundle. 

Such a divergence in their growth and dispersion can be also understood from the fact that as the bundle length $L_B$ grows to approach the end to end distance $L$, then $\alpha_f$ will approach $90^0$. Whereas, the angles in the bundle cannot exceed an angle given by $\tan^{-1}(\Omega r_i)$ because of normal contact forces between filaments. For example, in the ordered case of $n=2$, we have $r_i \approx d/2$, and the maximum $\phi_f$ can be at most $57.5^\circ$, and the system may accommodate high twist angles by pushing filaments further outward to form additional disordered layers.


\section{Applied Axial Torque}
\label{sec:torque}
\begin{figure}
\includegraphics[width=0.4\textwidth]{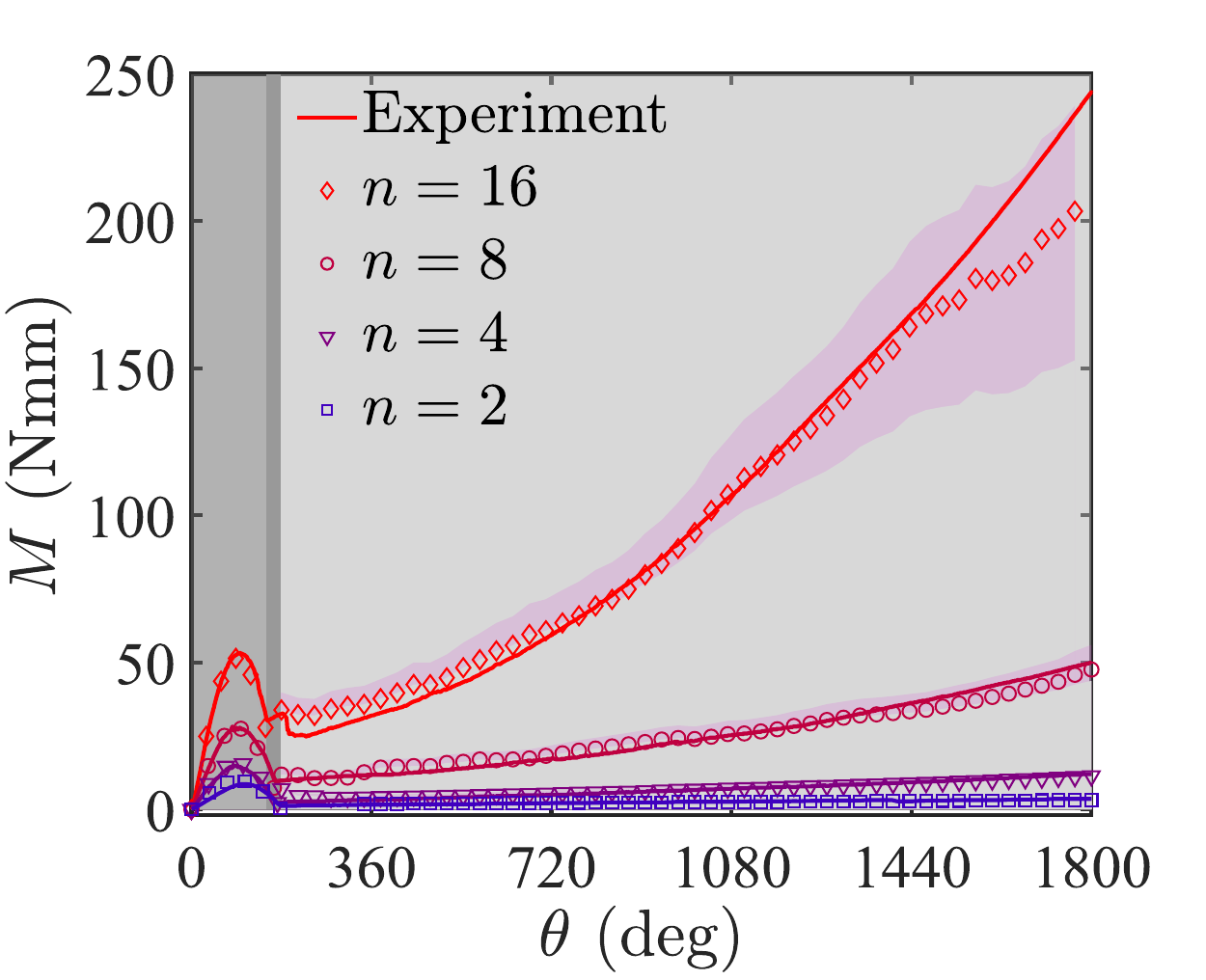}
	\caption{The comparison of estimated and measured torque $M$ as a function of $\theta$. 
    The estimated torque is calculated using Eq.~(\ref{eq:approximate_torque_at_bundle}). $\sum_{i=1}^n r_i=0.85nR_B$ (measured form optical image) is used, and $F_\mu = 1$\,N of friction is added to $F$ to mimic the effect of possible friction in the sliders.}
	\label{fig:tor}
\end{figure}

We now examine the consequences of the formation and evolution of the bundle on the torque needed as a function of applied twist. Figure~\ref{fig:tor} shows the applied torque $M$ measured as a function of $\theta$ with a Mark-10 digital torque gauge. A rapid increase and decrease is observed before the filaments come in contact. Such a peak has been noted previously in twisted filament pairs~\cite{Chopin2024}, where it was explained as arising as a consequence of the tension along the filaments pointing increasingly towards the axis of rotation as $\theta$ increases from $90^0$ to $180^0$.  

After contact, torque is observed to increase and this minima can be used to identify contact more clearly than in the plot of $L/L_0$ versus $\theta$, where a continuous change in its slope is observed. We plot $\theta_c$ obtained with this clear measurement in Fig.~\ref{fig:schem}, and note that it is well described by the estimate given by Eq.~(\ref{eq:thetac}). We observe that $M$ continues to rise increasingly non-linearly for various $n$. 

A small but rapid decrease in $M$ is also observed  after contact at an angle $\theta_r = 194^0$ in the case of $n=16$ before it increases non-linearly. From the analysis of the bundle structure, we inferred a rearrangement of the filaments relative to each other after coming in contact in a uniform circle at $\theta_c$. Consequently, we understand that the second decrease in $M$ occurs due to a radial instability analogous to a buckling instability observed in twisted ribbons~\cite{Chopin2022}, where the ribbons were observed to fold along their length leading to a decrease in bundle radius and torque. 

To relate the trends in $M$ to the evolving geometry of the filament structures, we note that each filament denoted with index $i$ applies a torque which depends on its tension $T_i$, the angle it makes with axis of rotation $\alpha_i$, and the distance of closest approach $r_i$ to the axis around which twist is applied. Accordingly, the analysis can be conducted under three conditions: (i) before contact, (ii) when the filaments come in contact in a circle with radius $R_c$, and (iii) when the filaments form a bundle. 
Accordingly, for $\theta \leq \theta_c$, 
\begin{equation}
    M(\theta) = \sum_{i=1}^n R_0 T_i  \cos{(\frac{\theta}{2})} \, {\sin\alpha_i}.
\end{equation}
Further, we assume the filaments are all equivalent $T_i = (F_1+F\mu/n)/\cos\alpha_i$ and $\alpha_i=\tan^{-1}(2R_0 \sin(\theta/2)/L)$.  
Substituting, we have
\begin{equation}
M(\theta) = (n{F_1}+F_\mu)  \frac{R_0^2}{L} \sin\theta. 
\label{eq:torbc}
\end{equation}
Ignoring further stretching of the filaments as they are twisted, trigonometry gives us
\begin{equation}
    {L} = \sqrt{\Big({L_0}^2 - 4R_0^2 \sin^2(\frac{\theta}{2})\Big)}\,,
    \label{eq:hyperboloid}
\end{equation}
where $L_0$ is the filament length before twist. 
For large enough $L/R_0$, the maximum would be at $\theta \approx 90^0$. 
We plot the estimated value with measured peak $M$ in Fig.~\ref{fig:tor} and observe good agreement. 

At contact, $\theta = \theta_c$, 
\begin{equation}
    M (\theta = \theta_c) = (n{F_1} + F_\mu)  \frac{R_0^2}{L} \sin\theta_c,
    \label{eq:torque_at_theta_C}
\end{equation}
with
\begin{equation}
    {L} = \sqrt{\Big({L_0}^2 - 4R_0^2 \sin^2(\frac{\theta_c}{2})\Big)}\,.
    \label{eq:hyperboloid}
\end{equation}
We compare the measured value at contact and the estimated value in Fig.~\ref{fig:tor}(c) and observe good agreement. 

After the formation of the bundle, i.e. $\theta > \theta_c$, 
\begin{equation}
    M(\theta) = \sum_{i=1}^n r_i T_i \sin{\alpha_i},
    \label{eq:Mlargetheta}
\end{equation}
where $r_i$ is the radial distance from the twist axis where the filament $i$ enters the bundle with
\begin{equation}
    \alpha_i = \tan^{-1}\big( \frac{2 \sqrt{R_0^2 - r_i^2}}{L-L_B} \big).
    \label{eq:inclination_at_triangular_part}
\end{equation}
If $T_i$ can be determined, then we can evaluate $M$ using Eq.~(\ref{eq:Mlargetheta}). 

Depending on the degree of stretching of each filaments $i$, force balance along the twist axis gives us 
\begin{equation}
F = \sum_{i=1}^n T_i\,\cos{\alpha_i},  
\end{equation}
where $T_i$ depends on the degree of stretching of the individual filament which can depend on their location relative to the twist axis. Only under conditions where the stretching is the same and angles are the same can it be assumed that $T_i = F_1/\cos\phi_f$. Those conditions are satisfied before the filaments touch and in symmetric bundles. 

Because we observe significant migration of the filaments which leads to an averaging effect on stretching, we assume that the stretching is roughly the same, and the tension in the filaments, $T_i = F_1/\cos\alpha_i$, we have 
\begin{equation}
M(\theta) =  F_1 \sum_{i=1}^n r_i \tan\alpha_i.
\label{eq:torque-approx}
\end{equation}

Substituting Eq.~(\ref{eq:inclination_at_triangular_part}) in Eq.~(\ref{eq:torque-approx}), we get
\begin{equation}
    M(\theta) = \frac{2F_1}{L-L_B}  \sum_{i=1}^n r_i \sqrt{R_0^2-r_i^2}.
\end{equation}
Now, for the tight bundle where $(r_i/R_0)^2 \ll 1$, 
\begin{equation}
M(\theta)\approx \frac{2R_0F_1}{L-L_B}  \sum_{i=1}^n r_i.
\label{eq:approximate_torque_at_bundle}
\end{equation}

Assuming $r_i$ distributions similar to those observed with x-ray CT, we find good agreement while comparing these estimates with data for $\theta = 720^0$ in Fig.~\ref{fig:tor}. This agreement illustrates the importance of the evolving geometry and axial load in determining the torque.  

\section{Energy Analysis}
\label{sec:model}

\begin{figure}
\includegraphics[width=0.45\textwidth]{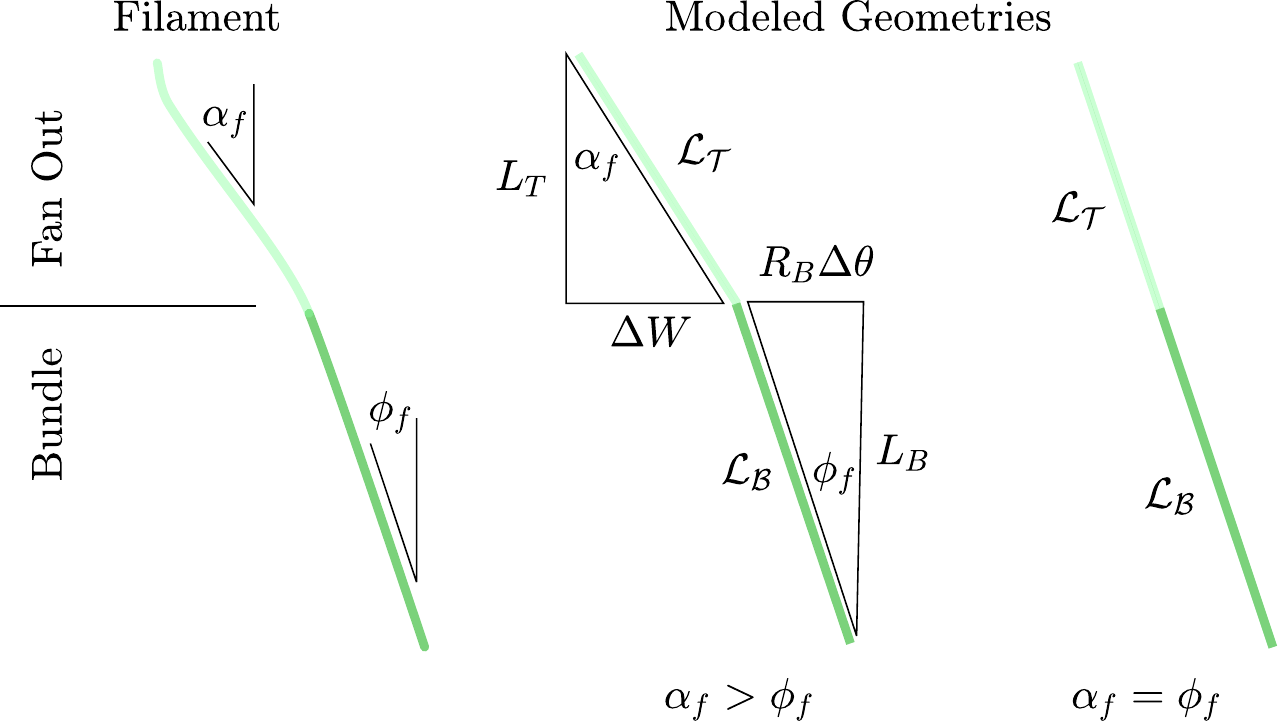}
	\caption{A schematic of the filament model where the shape of the twisted filament is projected onto a plane. The inclination angles in the bundle and fan out regions are different and depends on their distance from the twist axis. We assume the filaments to have the same angle of inclination in the bundle and the fan out region to simplify the analysis and understand the contribution of the various modes of deformation on the total elastic energy. 
 } 
	\label{fig:LB-model}
\end{figure} 

\begin{figure}
    \centering
   
    \includegraphics[width=1\linewidth]{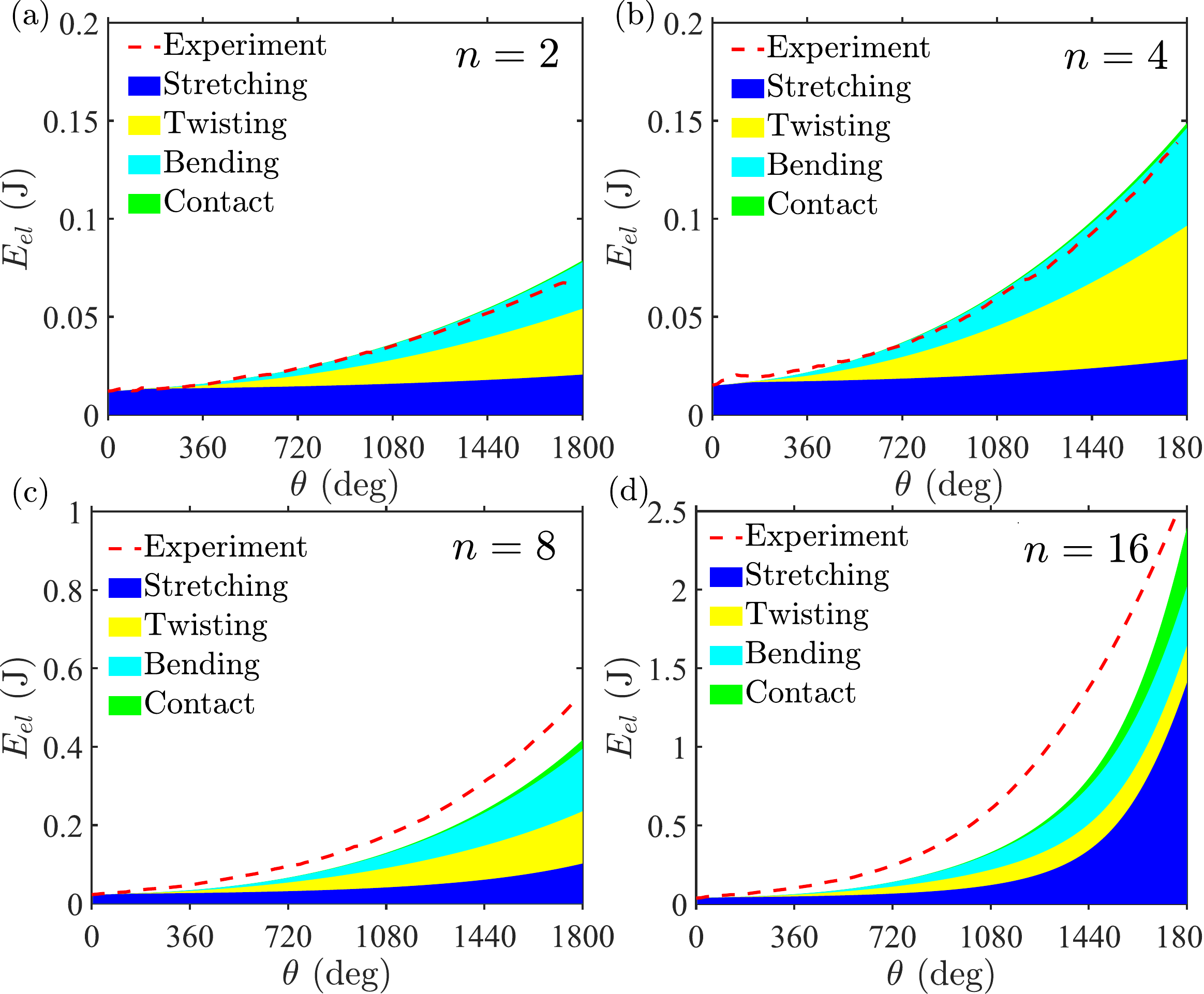}
    
    \caption{The estimated contribution due to filament stretching, bending, twisting and contact to the total stored energy using the elastogeometric model. The contribution of bending and stretching to the total stored energy increase rapidly. The increase of total energy matches the overall increasing trend observed in the experiments ($R_B=1.1$mm, $1.2$mm, $3$mm, and $5$mm respectively for $n=2, 4, 8,$ and $16$.)}
    \label{fig:energy}
\end{figure}

\begin{figure}
    \centering
    \includegraphics[width=1\linewidth]{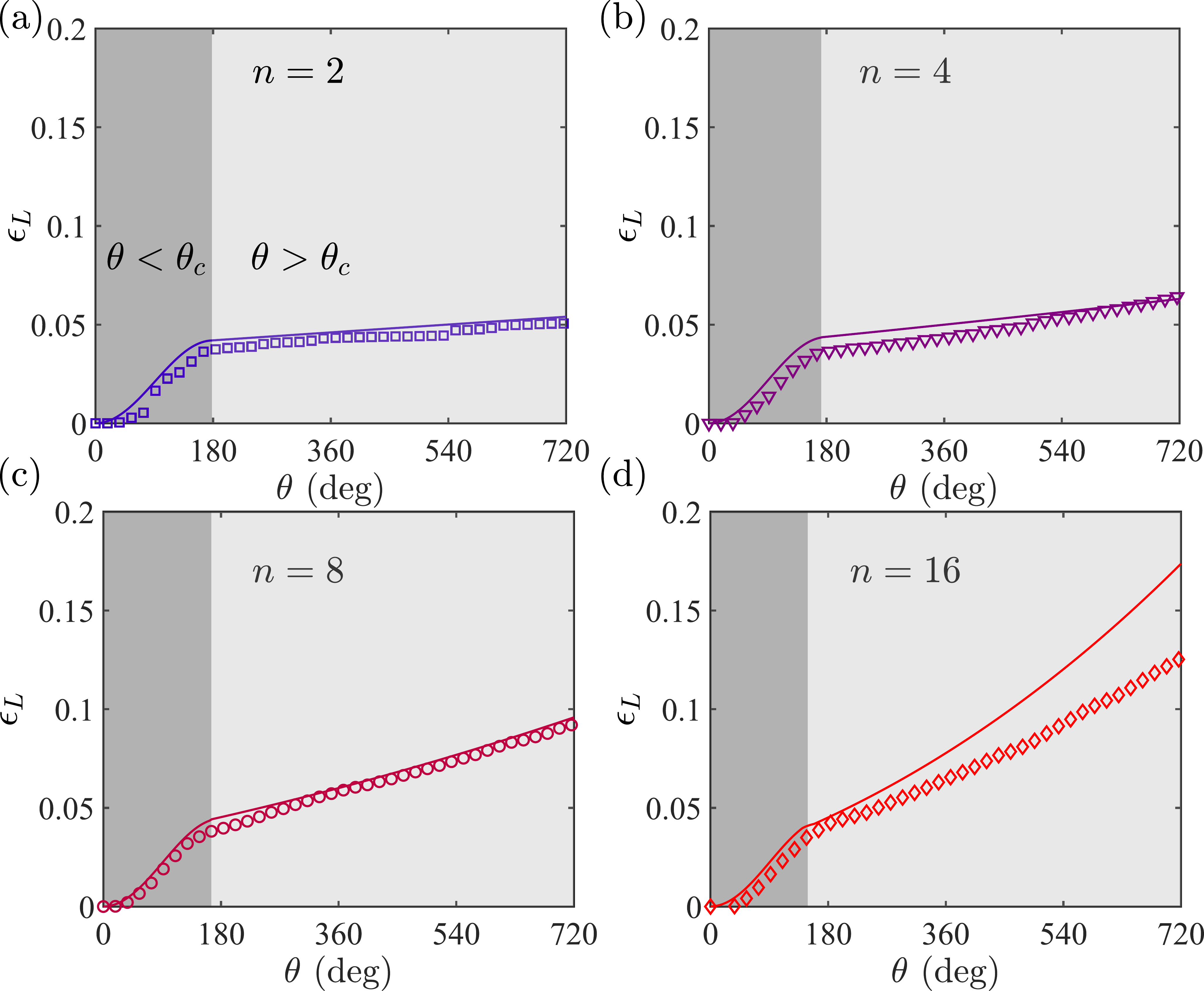}
    \caption{(a-d) The comparison of contraction length $ \epsilon_L = 1 - L/L_0$ versus $\theta$ with elastogeometric model for various $n$. Both the rapid increase before contact and crossover to slower contraction with lower slopes is captured by the model in the case of $n=2, 4$ and 8. The calculated trend for $n=16$ is systematically higher for $n=16$ because of the breakdown of the assumption that filament angles are equal in the bundle and fan out region. }
    \label{fig:contract}
\end{figure}

\begin{figure}
\includegraphics[width=1\linewidth]{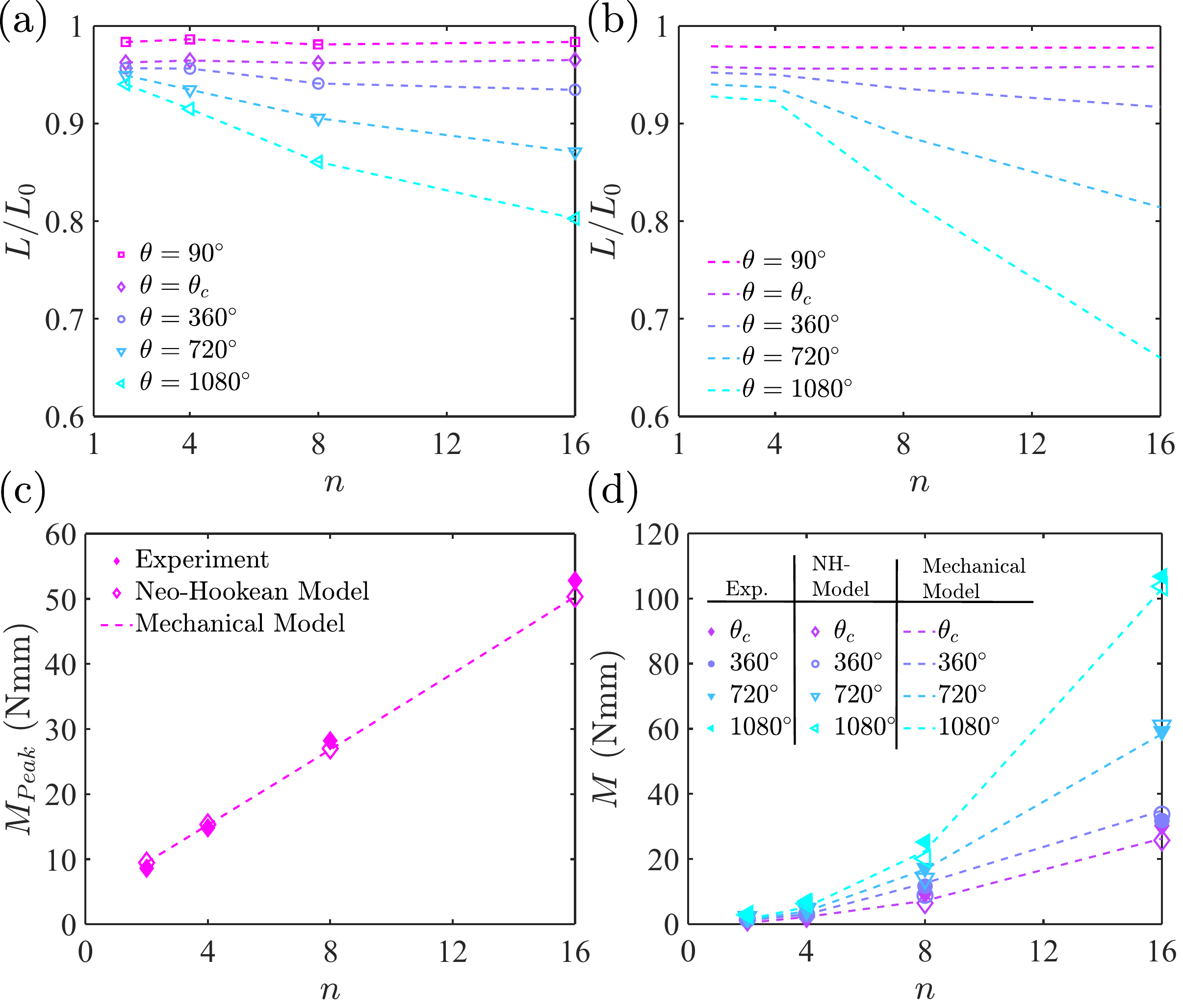}
\caption{(a,b) Measured and calculated end to end length as a function of number of filaments at different angles of twist. Deviations grow with increasing $\theta$ because of the approximations made in the model. (c) The measured and calculated peak torque versus number of filaments. (d) The measured and calculated torque versus number of filaments with increasing twist angle. The observed non-linear increase is well captured by the mechanical model and neo-Hookean analysis.}
\label{fig:torq-model}
\end{figure}

We next use an energy minimization approach to understand the formation of the bundle considering the elasticity of the filaments and the energy stored as a function of applied twist. The total elastic energy $E_{el}$ of the bundle is given by  
\begin{equation}
    E_{el} = E_s+E_t+E_b+E_c,
    \label{eq:Eel}
\end{equation}
where $E_s$ is the stretching energy, $E_t$ the twisting energy, $E_b$ the bending energy, and $E_c$ the contact energy. We obtain the force and the torque by derivation of the total elastic energy with respect to $L$ and $\theta$, respectively:  
 \begin{equation}
    F = \frac{\partial E_{el}}{\partial L}\Bigr\rvert_{\theta},
\end{equation}
and
 \begin{equation}
    M = \frac{\partial E_{el}}{\partial \theta}\Bigr\rvert_{L}.
\end{equation}

The filaments themselves are modeled using a neo-Hookean constitutive law \cite{ogden1997non}.  A representative single filament is used to capture the overall shape of the filaments in the helical bundle and fan out sections and a mean-field approach is used to considered a simplified picture of the contact interactions between filaments.  The filament geometry is modeled in the bundle as a helix of radius $R_B$, length $L_B$, and curvilenear length $\cL_B =\sqrt{L_B^2+R_B^2\Delta \theta^2}$, and as a straight line of length $\sqrt{(L-L_B)^2+\Delta W^2}$ in the fan out region where $\Delta W = 2 \sqrt{R_0^2-R_B^2}$ and $\Delta \theta = \theta-\theta_c$.  Thus, the curvilinear length reads:
\begin{equation}
    \cL = \cL_B + \cL_T+\delta \cL,
    \label{eq:cL}
\end{equation}
where $\delta \cL$ incorporate the deviation from the modeled geometry. In the fan out region, the filament curvature is zero and  in the bundle, it is given by : 
\begin{equation}
    \kappa = \frac{1}{R_B} \frac{\tan^2 \phi_f}{1+\tan^2 \phi_f} + \delta \kappa,
    \label{eq:kappa}
\end{equation}
where $\tan \phi_f = R_B\,\Delta \theta / L_B$. The first term corresponds to the curvature of a helix of radius $R_B$ and the second term accounts for deviations of the helical geometry.

In our modeled geometry, the bundle length $L_B$ is a parameter that can be set by imposing a stationary condition on the total energy with respect to $L_B$,  i.e. $dE_{el}/dL_B = (dE_{el}/d\lambda )(d\lambda/dL_B) = 0$.  We find $d\lambda/dL_B = \left( L_B/\cL_B - L_T/\cL_T  +  1/L_f\,(d\delta \cL/dL_B)\right)$. Thus, the condition $dE_{el}/dL_B = 0$ reads,
\begin{equation}
    0 =T \left( \cos \alpha_f - \cos \phi_f + \frac{d\delta \cL}{d\L_B}\right),
    \label{eq:normalforcebalance}
\end{equation}
where $T=\frac{1}{L_f}\frac{dE_{el}}{d\lambda}$ is the tension along the filament and $\cos \alpha_f =\frac{L_T}{\cL_T}$ and $\cos \phi_f = \frac{L_B}{\cL_B}$. Eq.~\ref{eq:normalforcebalance} can be interpreted as a force balance along the filament.  

Experimentally, we found that $\phi_f = \alpha_f - \Delta \phi(\theta)$ 
where $0<\Delta \phi(\theta) < 20^{\circ}$ over the range of experimental parameters explored (see Fig.~\ref{fig:x_ray_inclination1}). Thus, from $\cos \alpha_f - \cos \phi_f \approx \Delta \phi \sin \alpha_f$, we find that the difference of inclination angle between the bundle and the fan out region is a measure of geometrical imperfection term, $\frac{d\delta \cL}{d \L_B} \approx \Delta \phi \sin \alpha_f$ which can be treated as a small correction until $\theta = 720^{\circ}$ as $\Delta \phi \sin \alpha_f \approx 1/10$ at most. As a comparison, for $\theta = 1800^{\circ}$, $\Delta \phi \sin \alpha_f \approx 1/3$. Thus, we take $\delta \cL=0$, hence  $\alpha_f = \phi_f$ (see Fig.~\ref{fig:LB-model}), and $\delta \kappa = 0$. We then have: 
\begin{equation}
    \cL = \sqrt{L^2+\cW^2},
    \label{eq:cL}
\end{equation}
and
\begin{equation}
    \cL_B = \cL \frac{R_B \Delta \theta}{\cW},
    \label{eq:LB}
\end{equation}
where $\cW$ is the horizontal projected curvilinear length. We can show that  
\begin{align}
    \cW &= 2 R_0 \sin \frac{\theta}{2},\,\, {\rm for }\, \theta \leq \theta_c, \\
    \cW &= 2 \sqrt{R_0^2-R_B^2} + R_B \Delta \theta,\,\, {\rm for }\, \theta > \theta_c,
\end{align}
where the bundle radius $R_B$ is a complicated function of the mechanical parameters and is reminiscent of the waist size $\chi$ in the case of a twisted sheet~\cite{Chopin2022}.

The stretching energy for uniaxially stretched filaments is given by:
\begin{equation}
    E_s = \frac{n}{2} \mu V \left(\lambda^2+\frac{2}{\lambda} -3\right).
\end{equation}
where $\lambda = \cL/L_f$ is the stretching factor and $\cL$ the curvilinear length of the filament. The twisting energy is given by 
\begin{equation}
    E_t = \frac{n}{4} \mu V \left(\frac{R_0 \theta}{L_f}\right)^2 \frac{1}{\lambda}.
\end{equation}
Here, we assume that the twist density is homogeneously distributed along the filament. The bending energy is given by 
\begin{equation}
    E_b = \frac{3n}{2}\mu V \left(\frac{R_0}{L_f}\right)^2 \kappa^2 \frac{\cL_B}{\lambda},    
    \label{eq:bendingenergy}
\end{equation}
where $\kappa$ the curvature, and $\cL_B$ the curvilinear length of the filament in the bundle. 

To evaluate the contact energy in the bundle, we use a Hertz contact distributed over the bundle length $\cL_B$ in $n_c$ contacts. For $n=2$, there is exactly $n_c=1$ contacts, for $n=4$, there is exactly $n_c =4$. For $n=8$, we find on average $n_c \approx 12$ and for $n=16$, $n_c \approx 22$. Then, the Hertz contact energy is given by
\begin{equation}
    E_c = \frac{\pi}{2} n_c\mu\, \delta R^2\, \cL_B,
    \label{eq:Ec}
\end{equation}
where $\delta R$ is the indentation between two filaments upon self-contact. At equilibrium, the Hertz force $dE_c/d(\delta R) = \pi \mu \,\delta R\,\cL_B$ is balanced by the inward normal force $T \kappa \cL_B$, yielding 
\begin{equation}
    \delta R = \frac{1}{\pi} \frac{T\kappa}{\mu},
    \label{eq:deltaR}
\end{equation}
Thus,
\begin{equation}
    E_c = \frac{n_c}{2\pi\mu} \, T^2\kappa^2 \cL_B,
    \label{eq:Ec2}
\end{equation}

Using Eqs.~\ref{eq:kappa}, \ref{eq:LB}, and \ref{eq:cL}, the curvature of a filament reads $\kappa = \frac{1}{R_B}\left(\frac{\cW}{\cL}\right)^2$. Now, the bending energy given in Eq.~\ref{eq:bendingenergy} can be expressed more conveniently as
\begin{equation}
    E_b = \frac{3n}{8} \mu V \left(\frac{R_0}{L_f}\right)^2\left(\frac{\cW}{L}\right)^3\frac{L_f\Delta \theta}{R_B}\frac{1}{\lambda}.
\end{equation}

We can obtain the total elastic energy stored by integrating the measured torque as a function of applied twist and accounting for work done due to displacement of the load under constant applied force conditions, i.e. 
\begin{equation}
    E_{el} - F \Delta L =  \sum M \Delta \theta  - \sum F (L_0 - L), 
    \label{eq:Mexpt}
\end{equation}
where second term on l.h.s is due to stretching due to loading unstretched filaments before applying twist, first term on r.h.s. is due to the applied twist,  and second term on r.h.s. is the work done due to displacement of the load. 
We plot the integrated energy corresponding to Eq.~(\ref{eq:Mexpt}) in Fig.~\ref{fig:energy} and compare with different contributions according to Eq.~(\ref{eq:Eel}). Good overall agreement is observed in the case of $n=2$ and 4. However, while the overall increasing nonlinear trend is captured in the case of $n=8$ and 16, systematic deviations are also observed. Given the number of assumptions made in terms of simplifying the bundle geometry to calculate $\lambda$, and assumptions that the filaments all contribute equally to the energy, the degree of agreement is reasonable.  

Now the normalized contraction length using Eq.~\ref{eq:cL} reads: 

\begin{equation}
   \epsilon_L=1- \frac{L}{L_0} = 1- \frac{L_f}{L_0}\sqrt{\lambda^2-\left ( \frac{\cW}{L_f}\right )^2}. 
\end{equation}

We compare $\epsilon_L$ with measured values in Fig.~\ref{fig:contract}, and find good description of the contraction before contact in the hyperbolic hyperboloid regime. Further, we observe that the overall change in slope after contact and the relative increase with $n$ is well captured by our model. However, in case of $n=16$, systematic deviations are observed which may be expected given the growing number of assumptions with $n$. 

Further, plotting the measured and calculated $L/L_0$ versus $n$ side by side in Fig.~\ref{fig:torq-model}(a) and Fig.~\ref{fig:torq-model}(b) respectively, we observe that the deviations grow not only with increasing $n$ but also with $\theta$. This is consistent with breakdown of the assumption that $\alpha_c \approx \phi_c$, and an improved model of the crossover region from the helical to the fan out region is needed to further improve the description with increasing twist. 
We compare the measured and calculated peak torque before contact versus $n$ in Fig.~\ref{fig:torq-model}(c), and find good agreement. Further comparing the measured and calculated torque at various increasing twist angles, we also find good agreement for the nonlinear increase in torque measured with increasing $n$ at least up to $\theta = 1080^0$.  Thus, even within this simplified geometric model of the structure of the filaments, we find good overall description of the measured torque trends. 

The energy-based approach used previously~\cite{Chopin2024} found that the filament inclination angles in twisted filament pairs have to be the same in the bundle and fan out regions to have equilibrium. It is possible that inter-filament friction leads to some increase in $\alpha_i$ relative to $\phi_i$. However, we added a lubricant between the filaments and found that it does not lead to a significant change in the bundle length. Changing the order of applied strain by twisting to filaments and then applying tension showed some hysteresis, but even in this case the angles remained different. On the other hand, when we increased tension, we did find that $\alpha_i$ and $\phi_i$ approach each other. Since tension leads to higher normal forces between the filaments, one may expect the effect of friction to persist if they are important. However, this observation appears to rule out that explanation for the source of growing discrepancy. It is possible that excluded volume effects play a role in the growing divergence and remains worthy of further investigation. 

\section{Conclusions}
In conclusion, by performing complementary optical and x-ray measurements we elucidated the formation of multifilament bundles as individual filaments are bought together and twisted. Guided by these observations we analyzed the measured torque required to form them, and developed an elastogeometric model to estimate the energy stored in the system considering the hyperelastic nature of the filaments. The overall trends in measured torque and contraction length with twist angle is observed. Good quantitative agreement is observed for sufficiently low twist angles, but systematic deviations are found to grow because of the simplified geometry of the bundle considered in calculating the energies. While advancing the understanding of the system and providing useful estimates of the energy stored and the mechanical nature of the system, our study nonetheless illustrates the complexity of filament bundling and the need for further experimental and theoretical investigation.

\section{Acknowledgments}
    
This work was supported under U.S. National Science Foundation grant DMR-2005090.


\end{document}